# Carbon nanostructures obtained by underwater arc discharge of graphite electrodes: Synthesis and characterization.

*Juan G. Darias González[1], Lorenzo Hernández Tabare[1], Daniel Codorniu Pujals[2], Victoria Herrera Palma[1], Augusto Maury Toledo[4], Olimpia Arias de Fuentes[3], José X. Sierra Trujillo[1], Armando Bermúdez Martínez[2], Luis F. Desdín García[1]*

[1] Centro de Aplicaciones Tecnológicas y Desarrollo Nuclear (CEADEN) La Habana, Cuba
[2] Instituto Superior de Tecnologías y Ciencias Aplicadas (InSTEC), La Habana, Cuba
[3] Instituto de Ciencia y Tecnología de los Materiales (IMRE), La Habana, Cuba
[4] Centro de Estudios Avanzados de Cuba (CEAC) La Habana, Cuba.

darias@ceaden.edu.cu

**Abstract**

In the present work, the application of the method of underwater arc discharge of graphite electrodes for obtaining several carbon nanostructures is described. The analysis of the obtained products by Transmission Electron Microscopy (TEM), Scanning Electron Microscopy (SEM), Raman spectroscopy, Atomic Force Microscopy (AFM) and X-Ray Diffraction (XRD) showed that the samples collected from the material floating on the water surface were composed mainly by polyhedral onion-like particles, while those taken from the precipitate were a mixture multiwalled nano-tubes, onion-like particles and other graphitic structures. The main features of the obtained nanostructures are discussed.

## 1. Introduction

Since the discovery of fullerenes [1], and carbon nanotubes [2], there have been intensive efforts to produce and characterize nano-structured carbon materials, with the main objective being to exploit their properties applications. Examples of nano-structured carbon include nanotubes, nano-onions and nano-cones. Transmission electron microscopy (TEM), Scanning electron microscopy (SEM), Raman spectroscopy, Atomic Force Microscopy (AFM) and X-Ray Diffraction (XRD) are generally used to characterize these structures.

In the last years, plasma generated in liquids has been shown to be a promising method for nanoparticles synthesis [3], It can be used to produce a wide variety of nanomaterials [4-8], and eliminates the aforesaid drawbacks. Although this method is simpler than the conventional ones, it still requires a careful control of the relevant parameters to ensure the desired purity and structure of the products.

In this method several factors determine the nanoparticles composition and shape [9]. In the discharge, gas bubbles are generated acting as synthesis reactors. The bubbles dimensions depend on the arc power value and determine the size, shape and structure of the synthesized nanoparticles [10, 11]. The purpose of the research is to synthesize and characterize carbon nano-onions and nanotubes obtained by arc discharge in aqueous solution.

## 2. Experiment

### 2.1. Setup for arc discharge in solution

The carbon nanostructures used in this work were produced by applying a DC voltage across two graphite electrodes submerged in water, generating an arc discharge between them. The discharge current and voltage was 30 A and 16–17 V, respectively.

The synthesis station is composed of a DC power supply (5), a ballast resistor (1), an electronic control and DAQ system (2), a micropositioning system (3), a reaction chamber (4) and a chilling unit (6) (Fig. 1.).

The electrodes were submerged to a depth of 8 cm in 1.2MΩ resistivity distilled water contained on a 2.5L stainless steel vessel. As electrodes, spectroscopic pure graphite rods (⌀cathode = 12 mm and ⌀anode ~ 6 mm) were used.

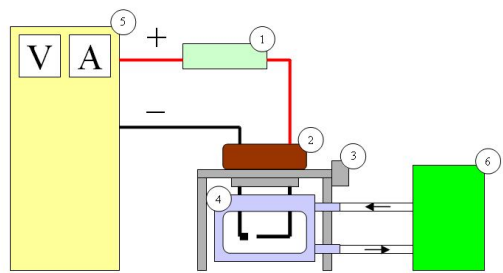

Figure 1: *Schematic diagram of the setup for synthesis of carbon nanoparticles.*

In our work a 5kW DC power supply (0-70V,0-70A) in its basic configuration was used (power transformer, diode bridge and capacitor bank). For arc stabilization a ballast resistor Rb ("positive" resistor) is placed in series with the electrodes.

The reaction chamber was made of stainless steel with two glass windows in order to have visual contact with the arc and the electrodes when aligning, testing and experimenting with the station.

The chamber has a double wall for cooling purpose. The bubbling forces the water to circulate inside the chamber thus homogenizing its temperature. The chilling unit forces the cold water to pass inside the double walled of the reaction chambers in order to cool the solution.

A shunt resistor is placed in the circuit in series with the electrodes. The voltage drop created in the shunt by the arc current is then amplified and filtered. A microcontroller (PIC18F4550) digitalizes the signal and comparing it with a prefixed value decides to move or not the anode toward the cathode by acting over the step motor.

## 2.2. Characterizations of carbon nanostructures

The Transmission Electron Microscopy (TEM) images were obtained using a Philips EM 208 microscope, with an accelerating voltage of 200kV.

The scanning electron microscope (SEM) images were obtained using a TESCAN MIRA 3 microscope.

The Atomic Force Microscopy (AFM) was carried out by the tapping method with a SPECTRA microscope from NT-MDT, using tips NSG-10, of monocrystalline Si doped with Sb.

Raman measurements were carried out with a microprobe setup (ISA-Jobin-Yvon, model Labram) equipped with a He–Ne laser (633 nm) and also using the Raman Spectrometer attached to the SPECTRA microscope cited above, with three different lasers (473, 633 and 785 nm).

The X-ray diffraction patterns were obtained in both RigakuDmax2100 and DRON4-M-10 diffractometers. Filtered Cu Kα radiation was used.

## 3. Discussion

In Figure 2 several images representatives of the samples taken from the material floating on the water surface after the arc discharge are showed.

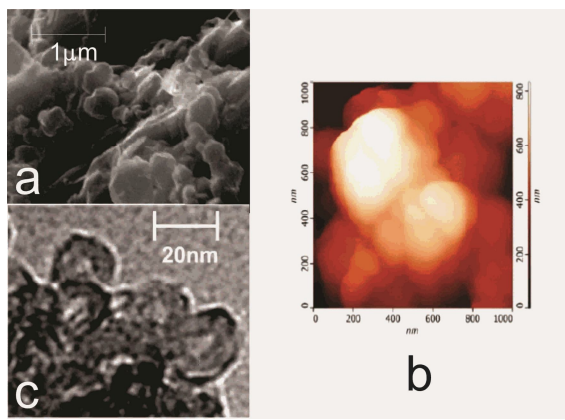

Figure 2: *Images of the samples taken from the material floating on the water surface after the arc discharge a) SEM, b) AFM, c) TEM.*

In the pictures of SEM (Fig. 2a) and AFM (Fig. 2b) it is possible to observe particles with linear dimensions of several hundred of nano-meters. The analysis by TEM (Fig. 2c) allows understanding that those particles are in fact agglomerates of polyhedral onion-like particles with average diameters in the range 10-50 nm. In the images of SEM other kind of particles with bigger size can also be observed. Following the reports of other authors [12], these particles may be associated with microscopic pieces of graphite separated from the electrodes and also with amorphous graphitic particles.

In figure 3a) a typical Raman spectrum of samples from the floating material is displayed. It is very similar to those reported for graphene and graphite [13], characterized by the presence of three main bands: the G-band linked to a Raman active optic in-plane stretching mode ($E_{2g}$), located near to 1580 cm-1; the D-band centered near to 1350 cm-1, whose presence in the spectra of sp2-nanocarbons is associated with the existence of defects in their crystalline lattice; and the G´ band, also called 2D by some authors, and located near of 2700 cm-1, corresponding to a two phonon (second-order) process.

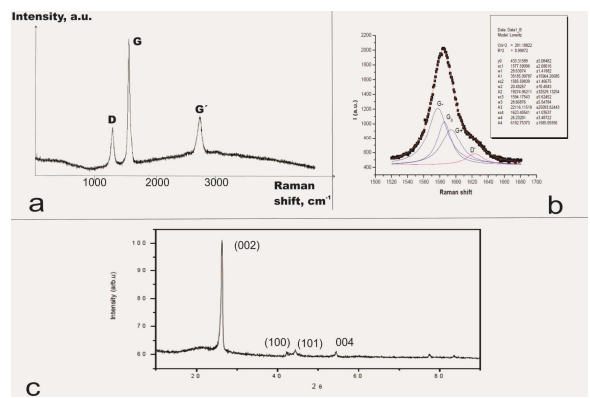

Figure 3: *Characterization of the samples taken from the material floating on the water surface. a) Raman spectrum, b) Deconvolution of the part of the Raman spectrum in the region 1520-1660 cm-1, c) X-ray diffraction pattern.*

A more detailed analysis of the Raman spectra in the neighbourhood of the G band shows that this band is composed by several peaks. The figure 3b) displays the deconvolution of the part of the spectrum between1520-1660 cm-1. There we can observe the well-known in the literature D´mode [14], that, like the D –band, is associated to some disorder in the lattice. The G band is well fitted by 3 lorentzian curves: one of them (Go) is located in the traditional position for this band in graphene or graphite (near 1580 cm-1), the second line (G-) is shifted to the left and the third (G+) is shifted to the right. The line Go can be linked with the graphite planes that composes the polyhedral onion-like particles. Other graphitic particles present in the samples can also contribute to this peak. The doublet (G+,G-) can be explained by the breaking of the degeneration in the center of the Brillouin zone of the optical in-plane transverse phonons (iTO) and the longitudinal optical (LO) phonon branches, due to the strain produced in the onion-like particles by the curvature of the graphene planes in some places. This phenomenon is similar to the one observed in the simple-walled carbon nanotubes (SWCNT), where the G band also splits because of the curvature [15], Then, the presence of the (G+, G-) is coherent with the presence of polyhedral onion-like particles.

In figure 3c) an X-Ray Diffraction pattern of the same group of samples is showed. There we can observe the relatively sharp and intense line of the graphite basal plane (002), located near 26º (2θ). The (100) and (101) lines in the angular interval 41 - 46º (2θ), and the (004) between 52 and 57º (2θ), are also observed, indicating a good crystalline order in these particles. The presence of an amorphous phase may also be noted in the low angles region.

A more detailed analysis of the (002) peak of the X-Ray Diffraction pattern allows to observe an evident asymmetry in this line. Taking into account the information obtained from TEM images and Raman spectroscopy, this asymmetry can be associated with the presence of different crystallographic zones in the polyhedral onion-like particles.

Samples of the precipitated material were also studied, using the already mentioned techniques. In the figure 4 several images of these samples are displayed. In this case, the typical picture observed by SEM and AFM is the coexistence of agglomerated particles, similar to those observed in the floating material, with elongated particles of different diameters (Fig. 4a and 4b).

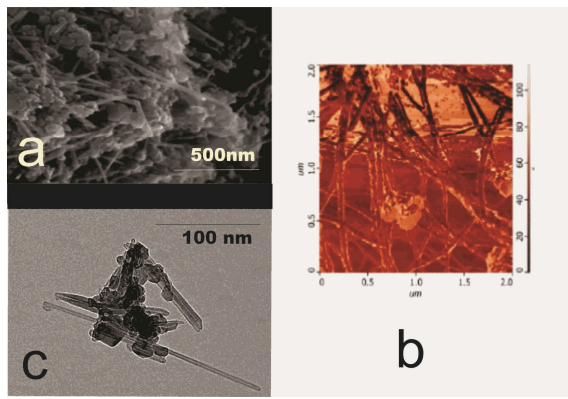

Figure4: *Images of the samples taken from the precipitate a) SEM, b) AFM, c) TEM.*

The figure 4c allows identifying the presence of clusters of onion-like particles together with multi-walled carbon nanotubes (MWCNT). This composition coincides with that reported by other authors [16], These nanotubes have lengths of 500 nm or more. In some cases they are grouped in bundles as it can be deduced from the analysis of the images of SEM and AFM.

It is worth to note that, although the main components of the samples taken from the floating material were onion-like particles, we also observe some MWCNT in most of them.

## 4. Conclusions

In summary, we investigated the controlled synthesis and characterization of carbon nano particles prepared via underwater arc discharge. The analysis of the obtained products by Transmission Electron Microscopy (TEM), Scanning Electron Microscopy (SEM), Raman spectroscopy, Atomic Force Microscopy (AFM) and X-Ray Diffraction (XRD) showed that the samples collected from the material floating on the water surface were composed mainly by polyhedral onion-like particles, while those taken from the precipitate were a mixture multiwalled nano-tubes, onion-like particles and other graphitic structures.